# Fast Polarization Switching Demonstration Using Crossed-Planar Undulator in a Seeded Free Electron Laser

Haixiao Deng, Tong Zhang, Lie Feng, Chao Feng, Bo Liu, Xingtao Wang, Taihe Lan, Guanglei Wang, Wenyan Zhang, Xiaoqing Liu, Jianhui Chen, Meng Zhang, Guoqiang Lin, Miao Zhang, Dong Wang and Zhentang Zhao*

*Shanghai Institute of Applied Physics, Chinese Academy of Sciences, Shanghai, 201800, P. R. China*

Fast polarization switching of light sources is required over a wide spectral range to investigate the symmetry of matter. In this Letter, we report the first experimental demonstration of the crossed-planar undulator technique at a seeded free-electron laser, which holds great promise for the full control and fast switching of the polarization of short-wavelength radiation. In the experiment, the polarization state of the coherent radiation at the $2^{nd}$ harmonic of the seed laser is switched successfully. The experiment results confirm the theory, and pave the way for applying the crossed-planar undulator technique for the seeded X-ray free electron lasers.

*Corresponding author: zhaozhentang@sinap.ac.cn

To satisfy the dramatically growing demand of photon sources in the biology, chemistry, physics and material sciences, high-intensity ultra-short radiation pulses with tunable wavelength are being routinely delivered to users by synchrotron radiation (SR) light sources and free-electron lasers (FELs) worldwide [1]. With the advent of X-ray FELs in the last ten years [2-5], a new era of X-ray science has arrived. Currently, the light source community is continuing to develop more sophisticated schemes in pursuit of e.g., fast polarization switching [5-17], full coherence [18-24], compact X-ray configurations [25-28] and multi-color operations [29-30].

Polarization describes the orientation of the light's electric field at a point in space over one oscillation period. In free space, the electric field of a polarized light may be oriented in a single direction, i.e., linear polarization, or it may rotate as the wave travels, i.e., circular or elliptical polarization. As is well-known, SR and FEL light sources usually involve a relativistic electron beam passing through a transverse periodic magnetic field, e. g., the undulator, and generating fully polarized radiation from the infra-red to hard X-ray regions. Because of the coupling between the radiation field and the electron motion in the undulator, the polarization state of the radiation is determined by the undulator magnetic field distribution. While a planar undulator only generates linearly polarized light, an elliptically polarized undulator (EPU) is capable for providing radiation pulses of arbitrary polarization.

Light with well-defined polarization is attractive for the experiments which aim at exploring the local symmetry of the samples, and has become a powerful tool for studying the electronic and magnetic properties of matter. To enhance the signal-to-noise ratio and improve the sensitivity to polarization weakly-dependent signals, one of the recent and growing features required by users is modulated fast polarization switching at rates from 10Hz up to 1kHz, e.g., between circular left-handed and right-handed, or between linear horizontal and vertical, or other combinations. However, the traditional light polarization control is accomplished by mechanical adjustment of the phase between the EPU magnets, a process of which lasts from one to many seconds depending on the magnitude of the desired changes in polarization. Therefore, multi-undulator techniques are developed for high repetition rate polarization switching: 1) Two canted EPUs deliver simultaneously two angularly distinct photon beams with different polarization states, and the fast polarization switching is accomplished by alternately selecting one of the two angularly-separated photon beams with a mechanical chopper [15]. 2) The electron beam is actively switched between two EPUs by a fast kick system, and thus capable for delivering one radiation pulse at a time with different polarization [16]. 3) Adjusting the phase between two successive EPUs changes the intensity of circular polarized radiations, and thus allows fast polarization switching from circular to linear in a hybrid undulator system [17]. In all the above-mentioned methods, the circularly polarized radiations are generated in EPU, and the fast switching can only be accomplished between two pre-set polarization handedness.

An alternative solution for arbitrarily polarized light is the scheme of crossed-planar undulator, which is originally proposed for SR [7] and FEL light sources [8]. It is based on the interference of horizontal and vertical radiation fields generated from two planar undulators in a crossed configuration, seen in Fig. 1. A pulsed electromagnetic phase-shifter between the crossed undulators is used to delay the electron beam and hence to control and switch the final polarization handedness. The advantage of the crossed-planar undulator is an arbitrary, programmable and fast polarization switching. A series of crossed-planar undulator experiments for incoherent SR sources have been carried out at BESSY [14]. Because of the coherent length of FEL radiations, various approaches using crossed-planar undulator for polarization control have been proposed for the modern FELs [8-13]. The key point of the crossed-planar undulator is to make the electromagnetic radiations along the horizontal and vertical axes as identical as possible, including the amplitude and the phase. For self-amplified spontaneous emission (SASE) FELs, the best time to place the crossed undulators is just before its power saturation, and then the density modulation induced by the SASE process will nearly remain the same while passing through the crossed undulators. However, the intrinsic spiky structure and the shot-to-shot fluctuation of SASE degrade the FEL polarization performance. While for the seeded FELs, the strong and controllable density modulation introduced by the interaction between the electron beam and the optical seed laser could result in stable and fully coherent output radiation. Thus with the crossed-planar undulator technique, excellent polarization control performances are expected in the seeded FELs.

In this Letter, we report the first experimental demonstration of fast polarization switching using crossed-planar undulator technique at the Shanghai deep ultraviolet free electron laser (SDUV-FEL) test facility at Shanghai Institute of Applied Physics. The experiment at SDUV-FEL [31] is a generic crossed-planar undulator proof-of-principle experiment that uses the 1047nm seed laser in the modulator. The fast polarization switching of the 523nm radiation, i.e., the $2^{nd}$ harmonic of the seed laser was successfully accomplished. The dependence of the polarization handedness as a function of the phase between the two adjacent undulators was measured and found to be in good agreement with theory, which indicates that scaling crossed-planar undulator to the seeded X-ray FELs may be possible.

The design considerations, commissioning, and the experimental details of fast polarization switching at SDUV-FEL are discussed in Refs. [12-13]. In the experiment, the linear accelerator provides a 145MeV electron beam with a bunch charge of 150pC, a normalized emittance less than 4mm-mrad, a projected energy spread of 0.2%, and a pulse duration of about 8ps FWHM without compression. The layout of the polarization switching experiment is schematically shown in Fig. 1. The main components consist of a electromagnetic planar modulator (EMU), a dispersive chicane (DS) and a crossed-planar radiator system (PMU-H, PS, PMU-V), several YAG/OTR screens for the electron and laser beam position and size measurements, correctors for beam trajectory control, and quadrupoles for beam matching and focusing. EMU (10 periods, period length of 65mm, and variable undulator magnetic field) and PMU-H (10 periods, period length of 50mm, and alterable undulator gap) was previously served as the twin-modulator in an echo-enabled harmonic generation FEL lasing experiment at SDUV-FEL [32]. In order to conduct the fast polarization switching experiment, a small electromagnetic phase-shifter PS with maximum $R_{56}$ of 5μm, a vertical planar undulator PMU-V (10 periods, period length of 50mm, and alterable undulator gap) and an optical bench for FEL radiation measurements were added at the downstream of PMU-V.

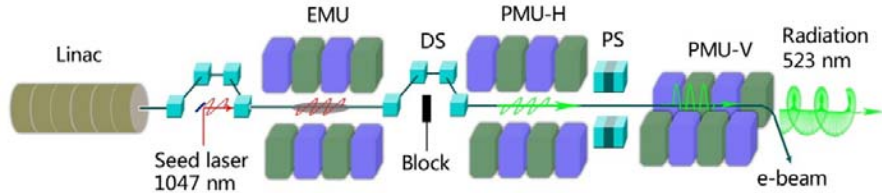

FIG. 1: (color online) Schematic of the polarization switching experiment at SDUV-FEL.

The seed laser is from the 1047nm drive laser (8.7ps FWHM pulse length), with a tunable pulse energy of 0-100μJ. Once the electron and laser beam overlapped spatially and temporally in the modulator (EMU) and the strength of the DS was optimized ($R_{56}$=3mm in the experiment), the micro-bunched electron beam is sent through the crossed-planar radiators which are tuned to be resonant at the $2^{nd}$ harmonic of the seed laser, to produce coherent FEL radiations at the central wavelength of 523nm. Using the modified optical replica method [33-34], critical parameters for the density modulation, such as the sliced beam energy spread and the energy modulation amplitude induced by the seed laser were precisely quantified. In this experiment, the measured sliced beam energy spread and the energy modulation amplitude is approximately 1.2 and 12keV, respectively, which corresponds to a strong bunching factor over 40% for 523nm radiation at the entrance of the radiators. Since the phase-shifter dispersion is pretty small and both the crossed radiators are only 10 periods, the electron beam can be treated as a rigid micro-bunched beam. Therefore, PMU-H and PMU-V should deliver identical radiations with nearly equivalent power and phase, but mutually perpendicular linear polarization. Then the phase-shifter PS can be used to fine tune the path difference between the horizontal and vertical polarized FELs, hence realize fast polarization switching of the combined radiation.

The polarization of the FEL light was characterized by a home-made division-of-amplitude photopolarimeter (DOAP), which consists of 1 optical lens, 3 beam splitters, 4 polarizers, 1 quarter-wave plate, and 4 photo-detectors. The incident light is divided into 4 separate beams, and the Stokes parameters ($S_0$, $S_1$, $S_2$, $S_3$) are measured simultaneously, thus allows to fully determine the polarization of the lights emitted from the crossed-planar radiators. The DOAP was intensively calibrated by the frequency-doubled (523nm) Nd:YLF laser. Here we recall that $S_1$ represents the fraction of intensity linearly polarized; $S_2$ is similar to $S_1$ but in a reference system rotated by 45°; $S_3$ is the fraction of intensity showing circular polarization. In particular, one finds $S_1/S_0=\pm1$ when the polarization is fully linear horizontal (vertical), while $S_3/S_0=\pm1$ for right-handed (left-handed) circular polarization. In order to test the setup, the seed laser was turned off and the PMU-V gap was changed from the maximum to 35mm (523nm resonance) while the PMU-H keeps resonance on 523nm, then a set of measurements of the combined spontaneous emission was performed. As the normalized Stokes' parameters $S_1/S_0$, $S_2/S_0$ and $S_3/S_0$ plotted in Fig. 2, when the undulator gap of PMU-V was fully opened, the light was mainly emitted from the PMU-H and completely horizontal polarized. As the vertical field component from the PMU-V was enhanced, the linear polarization degree of the combined spontaneous radiation decreases monotonously. For the case that both crossed-planar radiator are set to be resonant at 523nm, the combined spontaneous emission is completely un-polarized, even they were perfectly synchronized in spatial and temporal. This expected behavior depends on the fact that the coupling between the random phases of the vertical and horizontal spontaneous emission results in an arbitrary orientation of the combined light's electric field. This test actually presents a crossed-planar undulator experiment result on the storage ring light sources without monochromator, which demonstrates that the longitudinal coherence of the crossed lights is important and essential for the crossed-planar undulator technique.

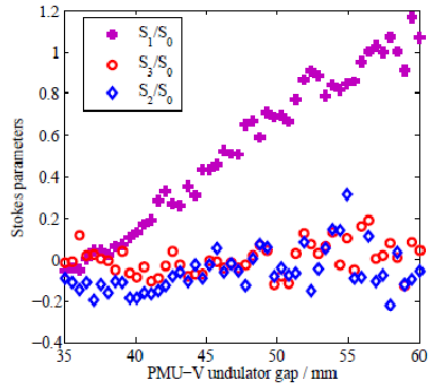

FIG. 2: (color online) The polarization performance of the combined spontaneous emission of crossed-planar undulator, characterized by the home-made DOAP system.

After optimizing all the resonant conditions of the undulators and the beam orbit in the polarization switching beam line, the seed laser was turned on and the coherent 523nm radiations generated from the crossed-planar undulator were reflected out of the vacuum chamber by a OTR screen and guided into an Ocean QE65000 spectrometer which is capable of detecting the signal from 350 to 1100nm. By kicking away the electron beam after PMU-H or opening up the gap of PMU-V, the coherent radiation generated in the two undulators respectively was sent into the spectrometers separately. Since the electron beam was accelerated on the crest in the linear accelerator, the linear beam energy chirp was minimized, thus both the vertical and horizontal polarized FELs were centered on 523nm wavelength, as drawn in Fig. 2, which shows that the two crossed lights with the same frequency are ready for effective combination and polarization switching. One should note that the spectra bandwidths broader than that in the Fourier-Transform-Limited case is due to the large residual energy curvature from the varying RF phase along the 8ps electron bunch and the limited spectrometer resolution of 2.4nm.

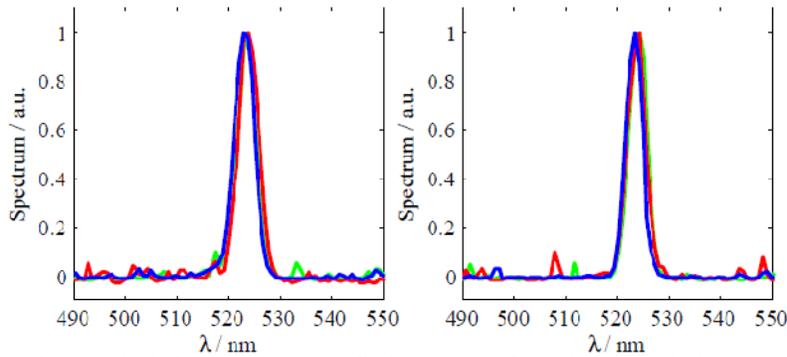

FIG. 3: (color online) Three typical shots of 523nm radiation spectra from the crossed-planar undulator. (a) represents the radiation with horizontally linear polarization generated in PMU-H. (b) represents the radiation with vertically linear polarization generated in PMU-V.

Then we tried to switch and characterize the polarization of coherent 523nm radiation by adjusting the phase-shifter. With a relatively stable beam and an optimized experimental condition, the measured Stokes parameters were given in Fig. 4. The circular polarization of the combined FEL was increased from 20% to 80%, while the linear polarization ($S_2$) was reduced from 95% to 20%. As a consequence, the combined FEL remained almost fully polarized during tuning. When we further increased the phase-shifter strength, the polarization switching between linear and circular was reproduced reasonably.

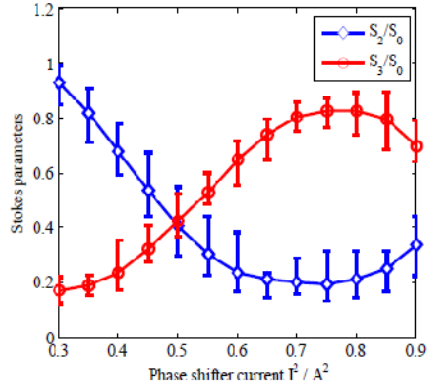

FIG. 4: (color online) The normalized Stokes parameters of coherent 523nm radiation measured at SDUV-FEL, where the phase-shifter current square is tuned from 0.3 to 0.9, corresponding to a phase delay about $0.6\pi$.

Generally speaking, one of the most key points for achieving perfect polarization switching is that the radiations from the two crossed undulators should be as identical as possible. In the experiment, we can slightly tune the gaps of crossed undulator and the quadrupoles to obtain almost equivalent FEL powers, while the radiation phase is determined by the seed laser. The other important point is that the radiations should be temporally and spatially overlapped. Because of the FEL slippage effects in the crossed-planar undulator system, the temporal overlap of 523nm radiations from PMU-H and PMU-V is naturally imperfect. It has been pointed out by start-to-end simulations that, the imperfect temporal overlap together with the inevitable residual beam energy chirp at SDUV-FEL will degrade the polarization performance of the combined FEL radiations by a factor of 10% [12]. Fig. 5 shows two typical transverse profiles of the radiations captured on a charge-coupled device (CCD). In order to clearly illustrate the effects of the spatial overlap, simulations using the two profiles were carried out. It is found that the polarization degree approaching 90% when the two radiations are optimally overlapped. Meanwhile, it decreases to 50% when these two radiations are separated by 200μm. The spatial overlap is determined by the beam orbit and envelope within the undulators, and the propagation in downstream free space. In the experiment, a shot-to-shot jitter of the photon beam position was clearly observed on CCD camera. The large polarization fluctuations in Fig. 4 was mainly contributed by the beam instability at SDUV-FEL due to the lack of feedback system.

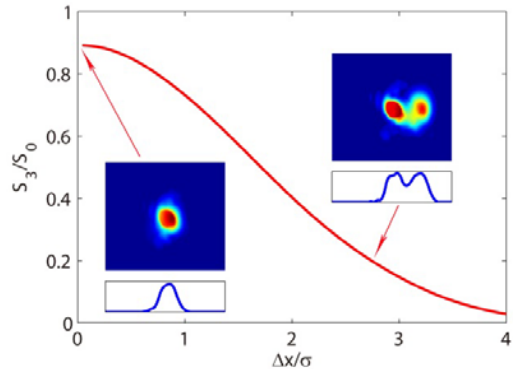

FIG. 5 (color online) Simulated polarization performance of the combined 523nm radiations vs the transverse position discrepancy of the two crossed radiations, where the experimentally observed transverse profiles are used.

In summary, using a seeded FEL configuration at SDUV-FEL, we presented the first experimental demonstration of fast polarization switching by crossed-planar undulator. The good agreement between the theory and our experimental results confirms the physics behind this technique. It can be easily extended to the existing [5] and future seeded FEL facilities [35-36] aimed from EUV to soft X-ray spectral region, in which fully coherent radiations with fast polarization switching are expected to open up new scientific opportunities in various research fields. It is worth stressing that, in a short-wavelength FEL, the polarization degradation due to the electron beam instability and radiation diffraction would be much more improved in comparison with current experiment carried out at visible light. Moreover, in this proof-of-principle experiment, the maximum repetition rate of the linear accelerator is 2Hz, while the greatest advantage of the crossed-planar undulator for fast polarization switching will definitely benefit the future high repetition rate FELs.

The authors would like to thank A. Chao, Z. Dai, Y. Ding, B. Faatz, H. Geng, Z. Huang, Q. Jia, Y. Li for helpful discussions, and SDUV-FEL team for FEL physics discussions, enthusiastic comments, and commissioning assistance. This work was partially supported by the Major State Basic Research Development Program of China (2011CB808300) and the National Natural Science Foundation of China (11175240, 11205234 and 11322550).